\documentclass{jpp}

\usepackage{graphicx}
\usepackage{natbib}

\ifCUPmtlplainloaded \else
  \checkfont{eurm10}
  \iffontfound
    \IfFileExists{upmath.sty}
      {\typeout{^^JFound AMS Euler Roman fonts on the system,
                   using the 'upmath' package.^^J}%
       \usepackage{upmath}}
      {\typeout{^^JFound AMS Euler Roman fonts on the system, but you
                   dont seem to have the}%
       \typeout{'upmath' package installed. JPP.cls can take advantage
                 of these fonts, if you use 'upmath' package.^^J}%
      }
  \else
  \fi
\fi

\ifCUPmtlplainloaded \else
  \checkfont{msam10}
  \iffontfound
    \IfFileExists{amssymb.sty}
      {\typeout{^^JFound AMS Symbol fonts on the system, using the
                'amssymb' package.^^J}%
       \usepackage{amssymb}%
       \let\le=\leqslant  
       \let\ge=\geqslant  
      }{}
  \fi
\fi

\ifCUPmtlplainloaded \else
  \IfFileExists{amsbsy.sty}
    {\typeout{^^JFound the 'amsbsy' package on the system, using it.^^J}%
     \usepackage{amsbsy}}
    {}
\fi

\newsavebox{\astrutbox}
\sbox{\astrutbox}{\rule[-5pt]{0pt}{20pt}}

\def\grad{{\rm grad}\,}

\def\<{\langle}
\def\>{\rangle}

\def\half{{\textstyle{1\over2}}}

\def\bi{\bf}

\def\rmi{i}
\def\rme{e}

\def\be{\begin{equation}}
\def\ee{\end{equation}}
\def\bea{\begin{eqnarray}}
\def\eea{\end{eqnarray}}
\def\nn{\nonumber}
\newcommand{\ms}{\noalign{\vspace{3pt plus2pt minus1pt}}}

\newfont{\myfont}{cmmib10}

\newcommand{\bphi}{\hbox{\myfont \symbol{30} }}

\newcommand{\bomega}{\hbox{\myfont \symbol{33} }}

\def\aa {{\it  Astron.\ Astrophys}. }

\def\apss{{\it  Astrophys.\ Space Sci}.}

\def\araa{{\it  Ann.\ Rev.\ Astron.\ Astrophys}. }
\def\apj{{\it  Astrophys.\ J}. }
\def\apjs{{\it  Astrophys.\ J.\ Suppl}. }
\def\aplett{{\it Astrophys.\ Lett}. }

\def\ajp{{\it  Aust.\ J.\ Phys}. }

\def\jgr{{\it  J.\ Geophys.\ Res}.\ }

\def\jpp{{\it  J.\ Plasma Phys}.}
\def\MNRAS{{\it  Mon.\ Not.\ R.\ Astron.\ Soc}. }

\def\nat{{\it  Nature} }

\def\prev{{\it  Phys.\ Rev}. }

\def\preve{{\it Phys.\ Rev.\ E} }
\def\prl{{\it  Phys.\ Rev.\ Lett}. }

\def\pasj{{\it  Publ.\ Astron.\ Soc.\ Japan} }

\title[Pulsar Electrodynamics]{Pulsar Electrodynamics: an unsolved problem}

\author[D. B. Melrose and R. Yuen]%
{D.\ns B.\ns M\ls E\ls L\ls R\ls O\ls S\ls E$^1$%
  \thanks{Email address for correspondence: donald.melrose@sydney.edu.au},\ns
R.\ns Y\ls U\ls E\ls N$^{1,2}$}
\affiliation{$^1$SIfA, School of Physics, The University of Sydney, NSW 2006, Australia\\[\affilskip]
$^2$Xinjiang Astronomical Observatory, Chinese Academy of Sciences, 40-5 South Beijing Road, Urumqi, Xinjiang, 830011, China}

\pubyear{2016}
\volume{Special Collection ``Plasma Physics of Gamma Ray Emission from Pulsars and their Nebulae''}
\date{: to be published in  Special Collection ``Plasma Physics of Gamma Ray Emission from Pulsars and their Nebulae''}

\begin{document}

\maketitle

\begin{abstract}
Pulsar electrodynamics is reviewed emphasizing the role of the inductive electric field in an oblique rotator and the incomplete screening of its parallel component by charges, leaving `gaps' with $E_\parallel\ne0$. The response of the plasma leads to a self-consistent electric field that complements the inductive electric field with a potential field leading to an electric drift and a polarization current associated with the total field. The electrodynamic models determine the charge density, $\rho$, and the current density, ${\bi J}$; charge starvation refers to situations where the plasma cannot supply $\rho$, resulting in a gap and associated particle acceleration and pair creation. It is pointed out that a form of current starvation also occurs implying a new class of gaps. The properties of gaps are discussed, emphasizing that static models are unstable, the role of large-amplitude longitudinal waves, and the azimuthal dependence that arises across a gap in an oblique rotator. Wave dispersion in a pulsar plasma is reviewed briefly, emphasizing its role in radio emission. Pulsar radio emission mechanisms are reviewed, and it is suggested that the most plausible is a form of plasma emission.
\end{abstract}

\begin{PACS}
%Authors should not enter PACS codes directly on the manuscript, as these must be chosen during the online submission process and will then be added during the typesetting process (see http://www.aip.org/pacs/ for the full list of PACS codes)
\end{PACS}

\section{Introduction}

Pulsars are strongly magnetized, rapidly rotating neutron stars. Their pulsed radiation is emitted in a pencil beam that sweeps across the line of sight as the star rotates. The observed emission can extend from $<100\,$MHz to extreme gamma-ray frequencies. There is an enormous body of detailed observational data on individual pulsars, including over 2000 radio pulsars \citep{L08} with approximately 10\% of these observed at high energy \citep{Aetal10}. The basic parameters that are measured (for most pulsars) are the period, $P$, and its rate of increase, ${\dot P}$. A  vacuum dipole model (VDM, cf. \S\ref{sect:VDM}) is used to identify a characteristic magnetic field, $B_*\propto(P{\dot P})^{1/2}$, and a characteristic age, $P/2{\dot P}$. Based primarily on the value $B_*$, pulsars may be separated into three classes: normal pulsars with $B_*$ of order $10^9\,$T, recycled (or millisecond) pulsars with much smaller $B_*$ and magnetars with much larger $B_*$. Despite the wide range of parameters, the radio emission from all three classes is remarkably similar. For most pulsars the integrated (over many pulse periods) pulse profile is known, and the `pulse window' is identified as the fraction of a  rotational period within which radiation is observed. Single pulses are observed from a subset of radio pulsars, and these exhibit a rich variety of features: subpulses, micropulses, jumps between orthogonal polarizations, and so on. The integrated pulse profile is stable over relatively long times, with some pulsars jumping between two or more different profiles (confusingly referred to as modes). Individual pulses show strong pulse-to-pulse variability, implying a large variance about a quasi-stable mean. The pulse window increases with decreasing frequency, and this is interpreted in terms of a radius-to-frequency mapping, with the beam pattern consistent with emission along a dipolar magnetic flux tube as it broadens with height. 

In this paper, we attempt to answer the question: Why is there still no widely accepted theory for the pulsar radio emission mechanism? Despite enormous progress in understanding the electrodynamics of pulsars  \citep{Mi90,Betal93,Me99}, there remain important unsolved problems, including the radio emission mechanism. We comment on three different approaches to identifying the radio emission mechanism: the first approach is observational, the second approach is theoretical, based on pulsar electrodynamics, and the third approach, also theoretical, is based on the properties of radio emission mechanisms. 

First: the enormous body of observational data, especially on pulsar radio emission, can be summarized in various `rules' that describe the properties of the observed emission, and one might expect such rules to provide severe constraints on possible emission mechanisms. However, such an observationally based approach is complicated by the fact that there seems to be exceptions to every rule, leading to differences of opinion as to the emphasis to be given to various rules and to the exceptions to them. An example concerns the polarization of pulsar radio emission. Early observations suggested a simple rule: the polarization is predominantly linear with a characteristic S-shaped sweep in position angle through the pulse. This was interpreted in terms of a rotating vector model \citep{RC69}. However, this rule is now recognized as, at best,  an  oversimplification. Observations of individual pulses show that they can be highly elliptically polarized, with a large pulse-to-pulse variation \citep{MS00,J04,ES04}, requiring a statistical interpretation \citep{Metal06}. Even in cases where there is a steady average swing in the position angle, it can jump by 90$^\circ$ at specific phases, with the sign of the circular polarization  reversing. This last feature, referred to as orthogonally polarized modes \citep{Setal84,MS00,McK02}, is strongly indicative of propagation through a birefringent medium with elliptically polarized natural modes \citep{PL00,WLH10,BP12}. An implication is that important features of the observed polarization must be due to propagation through a birefringent magnetospheric plasma, cf. \S\ref{sect:cold}, greatly complicating use of the polarization characteristics to constrain the emission mechanism.

Second: one might hope that a detailed model for pulsar electrodynamics, defined here to mean a self-consistent model for the plasma and electromagnetic field in the pulsar magnetosphere and the pulsar wind, would identify plausible locations where the radio-emitting particles are accelerated. There is a widely accepted model in which the magnetosphere is populated by relativistic electron/positron pairs that are continuously being created in the inner magnetosphere ($r\ll r_L=Pc/2\pi$) and escaping through the light cylinder, at radius $r=r_L$, to form the wind ($r\gg r_L$). This occurs along `open' magnetic field lines that define polar-cap regions, with the `closed' magnetosphere defined by field lines confined to $r<r_L$. The pairs are created in regions, called `gaps', where the parallel electric field, $E_\parallel$, accelerates charges to extremely high energy, such that they emit gamma rays that decay into pairs in the superstrong magnetic field \citep{S71}. In this model it seems plausible that the radio emission should be related to the pair creation, and hence to the location of gaps. However, there is no consensus on the location of gaps: there are models for inner gaps, near the stellar surface, outer gaps near the light cylinder, and slot gaps, near the boundary between open and closed field lines \citep{BS10,YS12}. Moreover, there are alternative models in which the acceleration by $E_\parallel$ occurs in a current sheet \citep{coroniti,U03,WS07,CB14,PS14}, rather than a gap. There are also alternative electrodynamics models, including an electrosphere \citep{KM85}, rather than a polar-cap model, and an ion-proton plasma \citep{jones}, rather than a pair plasma. Pulsar electrodynamics does not give a clear identification of the source region of the radio emission, or of the specific radio emission mechanism. 

Third: the number of conceivable radio emission mechanisms for pulsars is relatively modest, and  by considering all possible emission mechanisms, one might hope to identify at least one that is consistent with the observations. Possible mechanisms include plasma-type emission, curvature emission, linear acceleration emission and anomalous Doppler emission. Each mechanism has its own characteristics, including typical frequency and polarization, and one might expect to identify the most favourable mechanism by comparing the predicted and observed characteristics. However, this is not straightforward due to uncertainties in the location of the source region, the plasma parameters there and the properties of the radio-emitting particles. No consensus on which of these (or some other) is the most plausible mechanism has emerged. Moreover, pulsar radio emission requires a `coherent' emission mechanism, requiring either emission by bunches or some form of plasma instability, and again opinions differ as to which of these applies in pulsar radio emission.

The source region for the radio emission, although not well determined, is widely assumed to be on open field lines well inside the light cylinder. We accept this assumption, and concentrate our discussion on the region $r\ll r_L$, where retarded effects and the modification of the dipolar magnetic field due to magnetospheric current can be neglected (to lowest order in $r/r_L$).

Pulsar electrodynamics is summarized in \S\ref{sect:global}: the dichotomy in the use of two early models (VDM and RMM) is pointed out, and some more recent approaches involving numerical modelling are discussed briefly. Whether or not  the plasma can supply  the charge density, $\rho$, and and current density, ${\bi J}$, required by electrodynamic models is discussed in \S\ref{sect:response}, where we discuss screening of the parallel component of the inductive electric field. Failure of the plasma response to meet the requirements of the electrodynamics on $\rho$ and ${\bi J}$ are referred to as charge and current starvation, respectively, and implies the need for regions with $E_\parallel\ne0$. Such `gaps' are discussed in \S\ref{sect:gaps}, where it is argued that regions with $E_\parallel\ne0$ are intrinsically time dependent, and are probably associated with large-amplitude longitudinal oscillations. Interpretation of the radio emission depends in part on the dispersive properties of the plasma, and wave dispersion in a pulsar plasma is discussed in \S\ref{sect:dispersion}. Several of the suggested pulsar radio emission mechanisms are reviewed critically in \S\ref{sect:emission}, allowing the reader to see why there is no consensus on one specific mechanism. The results are discussed and conclusions are summarized in \S\ref{sect:discussion}.

\section{Global models}
\label{sect:global}

Pulsar electrodynamics is concerned with the large-scale distributions of fields and plasmas in the magnetosphere of a strongly magnetized, rapidly rotating neutron star. Two early models for the electrodynamics are the vacuum dipole model (VDM) and the rotating magnetosphere model (RMM). Both the VDM and RMM pre-dated the discovery of pulsars, and already led to controversy \citep{D47,A50} decades before the discovery of pulsars. There remains a dichotomy between the VDM and the RMM, with one or other used selectively for different purposes. A more recent approach involves numerical modelling using force-free electrodynamics (FFE). Although the RMM and FFE are sometimes regarded as equivalent, the distinction is maintained here. The defining assumption in a RMM is that the plasma is corotating with the star, whereas the defining assumption in FFE is that the electromagnetic force is zero. (Neither assumption can be strictly correct.)

\subsection{Maxwell's equations}

All models for electrodynamics are based on Maxwell's equations. These can always be written in the form
\bea
{\bf\nabla}\times{\bi E}&=&-{\partial{\bi B}\over\partial t},
\qquad
{\bf\nabla}\cdot{\bi B}=0,
\nn
\\
\ms
{\bf\nabla}\times{\bi B}&=&\mu_0{\bi J}+{1\over c^2}{\partial{\bi E}\over\partial t},
\qquad
{\bf\nabla}\cdot{\bi E}={\rho\over\varepsilon_0}.
\label{Me2}
\eea
These equations imply the wave equation,
\be
\left({1\over c^2}{\partial^2\over\partial t^2}-\nabla^2\right){\bi E}=
-\mu_0{\partial{\bi J}\over\partial t}-{{\bf\nabla}\rho\over\varepsilon_0}.
\label{e1}
\ee
%and the continuity equation for energy,
%\be
%{\partial\over\partial t}
%\left({|{\bf B}|^2\over2\mu_0}+\half\varepsilon_0|{\bf E}|^2\right)+{\bf\nabla}\cdot{{\bf E}\times{\bf B}\over\mu_0}=-{\bf J}\cdot{\bf E},%=-J_\parallel E_\parallel-{\bf J}_\perp\cdot{\bf E}_\perp,
%\label{energy1}
%\ee
%where ${\bf E}\times{\bf B}/\mu_0$ is the Poynting vector. 
The Maxwell stress tensor is
\be
({\bi T}_{\rm EM})_{ij}=-\left({B^2\over2\mu_0}+{\varepsilon_0E^2\over2}\right)\delta_{ij}
+{1\over\mu_0}B_iB_j+\varepsilon_0E_iE_j,
\label{Ms1}
\ee
and its divergence, combined with Maxwell's equations, gives the electromagnetic force per unit volume,
\be
{\bi f}_{\rm EM}=\rho{\bi E}+{\bi J}\times{\bi B}+{\partial{\bi P}_{\rm EM}\over\partial t},
\label{Ms3}
\ee
where ${\bi P}_{\rm EM}={\bi E}\times{\bi B}/\mu_0c^2$ is the momentum density in the electromagnetic field.

\subsection{Vacuum dipole model (VDM)}
\label{sect:VDM}

In the VDM the star is assumed to be surrounded by a vacuum, implying $\rho=0$, ${\bi J}=0$. The magnetic field is assumed to be due to a magnetic moment, ${\bi m}(t)$, rotating with the same angular frequency, $\bomega$, as the star, $\omega=2\pi/P$. Early versions of the VDM \citep{D47,D55} assumed the dipole is at the centre of a conducting sphere, and then there is a surface charge on the sphere that leads to a quadrupolar potential field in the surrounding vacuum. The VDM was applied to neutron stars by \cite{P67,P68}.

A major advantage of the VDM is that the fields can be evaluated explicitly, using 
\be
{\bi A}={\mu_0\over4\pi}{\bf\nabla}\times\left[{{\bi m}(t-r/c)\over r}\right],
\qquad
{\bi B}={\bf\nabla}\times{\bi A}
\qquad
{\bi E}=-{\partial{\bi A}\over\partial t},
\qquad
{\partial{\bi m}\over\partial t}=\bomega\times{\bi m},
\label{VDM1}
\ee
where $r=|{\bi x}|$ is the radial distance from the centre. When retarded effects are neglected, $t-r/c\to t$, ${\bi B}$ is the 	dipolar' field, and ${\bi E}\propto1/r^2$ is the `inductive' electric field,
\be
{\bi B}_{\rm dip}={B_*R_*^3\over2r^3}(3{\hat{\bi m}}\cdot{\hat{\bi r}}\,{\hat{\bi r}}-{\hat{\bi m}}),
\qquad
B_*={2\mu_0m\over4\pi R_*^3},
\qquad
{\bi E}_{\rm ind}={\mu_0m\omega\over4\pi r^2}({\hat{\bi m}}\cdot{\hat{\bi r}}\,{\hat{\bomega}}-{\hat{\bomega}}\cdot{\hat{\bi r}}\,{\hat{\bi m}}),
\label{VDM2}
\ee 
where ${\hat{\bi r}}={\bi x}/r,{\hat{\bi m}},{\hat{\bomega}}$ are unit vectors and $B_*$ is the magnetic field at the pole on the stellar surface, $r=R_*$. The retarded terms give additional contributions $\propto1/r^2$ and $\propto1/r$ to ${\bi B}$ and $\propto1/r$ to ${\bi E}$. The terms $\propto1/r$ give a radially outward Poynting vector $\propto{\hat{\bi r}}/r^2$ for $r\to\infty$, and integrating over a sphere at infinity gives the power in magnetic dipole radiation,
\be
{\cal P}={\mu_0m^2\omega^4\sin^2\alpha\over12\pi c^3},
\qquad
\cos\alpha={\hat{\bi m}}\cdot{\hat{\bomega}},
\label{VDM3}
\ee
where $\alpha$ is the obliquity angle. 

In the application of the VDM to pulsars, the rate of loss of angular momentum, ${\cal P}/\omega$, to magnetic dipolar radiation, is equated to the rate of loss of rotational angular momentum, $I_*{\dot{\omega}}$, where $I_*$ is the moment of inertia of the star, which is assumed to be approximately the same for all neutron stars. Using (\ref{VDM2}) and (\ref{VDM3}), this determines the magnetic field $B_*\sin\alpha=6\times10^{15}(P{\dot P})^{1/2}\,$T in terms of the observables, where the constant of proportionality is uncertain by a factor of order unity. The age is identified as $P/2{\dot P}$ on the basis of the VDM with $\alpha$ assumed independent of time. The assumption that $\alpha$ is constant is artificial: the VDM implies that the torque due to magnetic dipole radiation has a component leading to the slowing down, and a component orthogonal to it that causes $\alpha$ to decrease on the same time scale as the slowing down \citep{DG70}. Observations suggest that alignment occurs, but only over approximately $10^7\,$yrs \citep{LM88}, inconsistent with the prediction. It is now accepted that the slowing down of a pulsar is due to a wind.

If plasma is present in the magnetosphere, at sufficiently low density such that it does not modify ${\bi E}$ significantly, the VDM implies an electric drift determined by ${\bi E}_{\rm ind}$:
\be
{\bi v}_{\rm ind}={{\bi E}_{\rm ind}\times{\bi B}\over B^2}=\omega r\,{\hat{\bi m}}\cdot{\hat{\bi r}}\,
{3{\hat{\bi m}}\cdot{\hat{\bi r}}\,{\hat{\bomega}}\times{\hat{\bi r}}
-3{\hat{\bomega}}\cdot{\hat{\bi r}}\,{\hat{\bi m}}\times{\hat{\bi r}}
-{\hat{\bomega}}\times{\hat{\bi m}}\over3({\hat{\bi m}}\cdot{\hat{\bi r}})^2+1}.
\label{vind}
\ee
This velocity is of the same order of magnitude, $\omega r$, as the corotation velocity, $\bomega\times{\bi x}$, but differs from it in both direction and magnitude. The velocity (\ref{vind}) corresponds to a rotating vector field, in the sense that the field pattern is periodic with period $P=2\pi/\omega$. The plasma flow is not periodic: a given blob of plasma does not return to the same location after one rotation.

\subsection{Rotating magnetosphere model (RMM)}
\label{sect:RMM}

Soon after the discovery of pulsars, a RMM was proposed \citep{GJ69}. The basic assumption is that the neutron star is surrounded by plasma that corotates with the star, analogous to planetary magnetospheres, like those of the Earth and Jupiter \citep{HB65}. For simplicity, \cite{GJ69} assumed alignment of the magnetic and rotation axes; subsequently the assumption $\sin\alpha=0$ dominated much of the pulsar literature on the RMM. This neglects some essential electrodynamics: an aligned model has no inductive electric field and no magnetic dipole radiation. An oblique version of the RMM \citep{HB65} predated the discovery of pulsars, and the discussion of the RMM in this paper presupposed the more general case with $\sin\alpha\ne0$. A notable distinction between RMMs for planetary and pulsar magnetospheres is that the distance, called the light cylinder, where the corotation speed equals the speed of light, is important in the pulsar case, where it separates the inner magnetosphere from the pulsar wind. In a RMM, the charge and current densities become infinite at the light cylinder \citep{GJ69}, and in the discussion of RMMs here we assume they only apply well within the light cylinder. 

Plasma motion in a (pulsar) magnetosphere is determined by the electric drift velocity, and the electric field corresponding to corotation is ${\bi E}_{\rm cor}=-{\bi v}_{\rm cor}\times{\bi B}$ with ${\bi v}_{\rm cor}=\bomega\times{\bi x}=\omega r{\hat{\bphi}}$. The electric drift implies only the component, ${\bi v}_{\rm cor\perp}={\bi v}_{\rm cor}-{\bi b}{\bi b}\cdot{\bi v}_{\rm cor}$, perpendicular to ${\bi B}=B{\bi b}$. \cite{HB65} noted that, in an obliquely rotating planetary magnetosphere, the parallel component, $v_{\rm cor\parallel}={\bi b}\cdot{\bi v}_{\rm cor}$, can be set up mechanically, due to a trapped particle being reflected from moving mirror points as the star rotates. This is not possible in a pulsar magnetosphere: due to the superstrong magnetic field, electrons and positrons radiate away their perpendicular energy extremely rapidly, so that their orbital magnetic moment, $mv_\perp^2/2B$, is identically zero. This precludes any mirroring. There is no other mechanical force that can set up or maintain $v_{\rm cor\parallel}$ in a neutron star magnetosphere. Hence `corotation' in an oblique pulsar should be interpreted as the electric drift velocity ${\bi v}_E={\bi v}_{\rm cor\perp}$ with
\be
{\bi v}_{\rm cor\perp}=\omega r{[3({\hat{\bi m}}\cdot{\hat{\bi r}})^2+1]{\hat{\bomega}}\times{\hat{\bi r}}
-{\hat{\bomega}}\times{\hat{\bi m}}\cdot{\hat{\bi r}}(3{\hat{\bi m}}\cdot{\hat{\bi r}}\,{\hat{\bi r}}-{\hat{\bi m}})
\over3({\hat{\bi m}}\cdot{\hat{\bi r}})^2+1}.
\label{RMM1}
\ee

The corotation electric field has both an inductive (divergence-free) and a potential (curl-free) component \citep{M67},
\be
{\bi E}_{\rm cor}={\bi E}_{\rm ind}+{\bi E}_{\rm pot},
\qquad
{\bi E}_{\rm ind}=-{\partial{\bi A}\over\partial t},
\qquad
{\bi E}_{\rm pot}=-{\bf\nabla}\Phi_{\rm cor},
\qquad
\Phi_{\rm cor}=\bomega\times{\bi x}\cdot{\bi A}.
\label{RMM2}
\ee
The inductive field is unchanged from its value in vacuo (in the VDM), and the additional potential field is due to the corotation charge density, $\rho=\rho_{\rm cor}$, with
\be
{\rho_{\rm cor}\over\varepsilon_0}={\bf\nabla}\cdot{\bi E}_{\rm cor}=-2\bomega\cdot{\bi B}
+(\bomega\times{\bi x})\cdot{\bf\nabla}\times{\bi B}.
\label{RMM3}
\ee
The final term in (\ref{RMM3}) may be rewritten using Amp\`ere's equation (\ref{Me2}). 

Amp\`ere's equation can be separated into three parts. One part with ${\bi J}=0$ corresponds to the VDM, with the retarded parts of ${\bf\nabla}\times{\bi B}$ and the displacement current, ${\bi J}_{\rm disp}=\varepsilon_0\partial{\bi E}/\partial t$, due to ${\bi E}_{\rm ind}$ in balance.  A second part involves the contribution to ${\bi J}_{\rm disp}$ from ${\bi E}_{\rm pot}$ being balanced by a corotation current density. The explicit form for ${\bi E}_{\rm pot}$ for $r\ll r_L$ is
\be
{\bi E}_{\rm pot}=-{\omega^2m\over4\pi r^2c^2}\,(
3{\hat{\bomega}}\cdot{\hat{\bi r}}\,{\hat{\bi m}}\cdot{\hat{\bi r}}\,{\hat{\bi r}}
-{\hat{\bomega}}\cdot{\hat{\bi m}}\,{\hat{\bi r}}
-{\hat{\bi m}}\cdot{\hat{\bi r}}\,{\hat{\bomega}}
-{\hat{\bomega}}\cdot{\hat{\bi r}}\,{\hat{\bi m}}),
\label{Epot}
\ee
implying \citep[see][]{FAS77}
\be
{\bi J}_{\rm cor}=-\varepsilon_0{\partial{\bi E}_{\rm pot}\over\partial t}=
{\omega^2m\over4\pi r^2c^2}\,[{\hat{\bomega}}\times{\hat{\bi m}}\cdot{\hat{\bi r}}\,
(3{\hat{\bomega}}\cdot{\hat{\bi r}}\,{\hat{\bi r}}-{\hat{\bomega}})-{\hat{\bomega}}\cdot{\hat{\bi r}}\,{\hat{\bomega}}\times{\hat{\bi m}}].
\label{RMM4}
\ee 
The remaining part involves additional currents in the plasma, including $\rho_{\rm cor}{\bi v}_{\rm cor\perp}$ and a remaining part, ${\bi J}_{\rm ext}$ say, that are balanced by an additional contribution to ${\bf\nabla}\times{\bi B}$. Then (\ref{RMM3}) implies
\be
\rho_{\rm cor}={\varepsilon_0\over1-|{\bi v}_{\rm cor\perp}|^2/c^2}\left[-2\bomega\cdot{\bi B}
+(\bomega\times{\bi x})\cdot\left(\mu_0{\bi J}_{\rm ext}+{1\over c^2}{\partial{\bi E}_{\rm ind}\over\partial t}\right)\right],
\label{RMM6}
\ee
which reduces to the Goldreich-Julian charge density in the aligned case with ${\bi J}_{\rm ext}=0$.

The plasma velocity (\ref{RMM1}) in the oblique corotation model may be regarded as the sum of the electric drift velocities due to ${\bi E}_{\rm ind}$ and ${\bi E}_{\rm pot}$. This allows one to identify a class of intermediate models \citep{MY12,MY14} with electric drift velocity
\be
{\bi v}'_E=y{\bf v}_{\rm ind}+(1-y){\bf v}_{\rm cor\perp},
\label{ymodel}
\ee
reducing to the VDM and the RMM for $y=1$ and $y=0$, respectively.

In early RMMs it was assumed that the stellar surface is a source of charge (`primary' charges), and that pair creation (`secondary' charges) in gaps is needed to allow the plasma to provide the required charge density. However, the assumptions that charges arise from the stellar surface leads to an `electrosphere' \citep{KM85,S04}, rather than the widely accepted polar-cap model. As discussed below, older electrostatic models for pair creation in gaps are unstable to temporal perturbations, and the acceleration by $E_\parallel\ne0$ is in intrinsically time-dependent structures. This has led to the suggestion, which we favour, that the stellar surface plays no significant role in populating the magnetosphere with charges.

\subsection{Force-free electrodynamics (FFE)}
\label{sect:FFE}

More recently, force-free electrodynamics (FFE) has been used as the basis for models for the global electrodynamics, particularly for the region $r\gtrsim r_L$ covering the transition from the inner magnetosphere to the pulsar wind \citep{FFE1,FFE2,LST12}. FFE may be interpreted as a modified form of magnetohydrodynamics (MHD) in which relativistic effects and the displacement current are included,  and the inertia of the plasma is neglected. As in MHD, the assumption $E_\parallel=0$ is made in FFE.

The force-free condition is ${\bi f}_{\rm EM}=0$ in (\ref{Ms3}), and this reduces to the standard form assumed in the FFE, $\rho{\bi E}+{\bi J}\times{\bi B}=0$, when the momentum density of the electromagnetic field is neglected. The plasma inertia is included implicitly through the polarization (or inertia) current, and neglecting the plasma inertia corresponds to omitting the polarization current from ${\bi J}$. This assumption may be justified in pulsar magnetospheric plasma by noting that the Alfv\'en speed, $v_A$, is much greater than the speed of light, and that the ratio of ${\bi J}_{\rm pol}$ to ${\bi J}_{\rm disp}$ is $c^2/v_A^2\ll1$. Stresses that are transmitted by Alfv\'en waves in MHD are transmitted at $v_0=v_A/(1+c^2/v_A^2)^{1/2}\approx c$ in a pulsar plasma. Although the force-free condition cannot be strictly valid, because the slowing-down torque must be transmitted from the wind to the star by the Maxwell stress in the magnetosphere, the non-zero ${\bi f}_{\rm EM}$ is of the same order as that in the VDM. It has been shown that an FFE model  effectively reduces to the VDM \citep{P12} as the vacuum limit is approached.

A major achievement of FFE models is in describing the transition region from the inner magnetosphere at $r\ll r_L$ to the wind zone at $r\gg r_L$. The FFE solutions break down, leading to discontinuities, near the light cylinder and along the last closed field lines \citep{coroniti,U03,CB14,PS14}.  The discontinuities are interpreted in terms of a $Y$-shaped current sheet, with the solutions on either side of the sheet linked by the boundary conditions at the sheet.  This was first recognized for an aligned rotator; solutions for an oblique rotator show similar features \citep{TSL13}.

The assumption $E_\parallel=0$, made in MHD and FFE, cannot apply everywhere in a pulsar magnetosphere. A realistic pulsar model needs to include regions with $E_\parallel\ne0$ \citep{Ketal12}, where acceleration and associated pair creation occur. The regions with $E_\parallel\ne0$ are assumed to be localized, as gaps in RMMs and current sheets in global FFE models. One interpretation of the need for regions with $E_\parallel\ne0$ is that (in MHD and FFE) $\rho$ and ${\bi J}$ are determined by the electrodynamics, without reference to whether or not the plasma can supply the required charges and currents, and a model based on $E_\parallel=0$ must be modified when the plasma cannot support the required $\rho$ or ${\bi J}$, in particular when these become infinite. %Even in an aligned model, the implicit assumption that the plasma can provide the required $\rho$ and ${\bi J}$ everywhere cannot be satisfied \citep{C05}. This problem is overcome in RMMs through the inclusion of gaps, where particle acceleration by  $E_\parallel\ne0$ and pair creation occur. In global models based on FFE, gaps are not included explicitly, and then the continuity of the models themselves break down at the specific surfaces identified as current sheets. 
%One suggestion is that such current sheets replace gaps as the regions where particle acceleration by $E_\parallel\ne0$ occurs. 
If the source of the radio emission is assumed to be close to the acceleration regions, the identification of the acceleration site with current sheets in a global FFE model would imply that the radio source is near the light cylinder. It seems more plausible that the radio source is at $r\ll r_L$, and that the electrons and positrons that produce the radio emission are accelerated in gaps at $r\ll r_L$.  One approach to modelliing the effects of such regions using FFE is to replace the assumption $E_\parallel=0$ in `ideal' FFE by allowing a non-zero resistivity \citep{Ketal12,LST12}. Another approach is to complement a global FFE model with PIC calculations describing localized pair creation \citep{T10,TA13,Cetal15,PSC15}. Although these approaches are encouraging, they have yet to result in a self-consistent model that is useful in constraining the location and properties of the radio source region.

\section{Response of a pulsar plasma}
\label{sect:response}

The response of the plasma %consists of the $\rho$ and  ${\bi J}$ induced in the plasma by the self-consistent electromagnetic field, which is to be determined by Maxwell's equations with  $\rho$ and ${\bi J}$ regarded as functions of this field. The response of a pulsar plasma 
needs to be considered in two separated contexts: the response to the `background' electromagnetic field, and wave dispersion in a pulsar plasma. The first of these is discussed in this Section, and the second is discussed in \S\ref{sect:dispersion}.

\subsection{Response to the background field}

In a pulsar magnetosphere, the `background' electromagnetic field is that due to the rotating magnetized star, approximated here by a rotating magnetic dipole at the centre of the star. Currents flowing in the magnetosphere provide an additional magnetic field. Here we concentrate on the region $r\ll r_L$, where the modification to the dipolar field as a result of magnetospheric currents (and due to retarded effects) is small, and can be neglected to a first approximation. 

The electrodynamic problem is to determine $\rho$, ${\bi J}$ and an associated potential field, ${\bi E}_{\rm pot}$, given the background electromagnetic field. In the VDM the solution is known:  $\rho$ and ${\bi J}$ are assumed zero, and ${\bi E}_{\rm pot}$ is attributed to a surface charge distribution on the star. In the RMM, ${\bi E}$ is, by hypothesis, the corotation field, ${\bi E}_{\rm cor}$, and $\rho$ and  ${\bi J}$ are determined from it by Maxwell's equations, cf.\ (\ref{RMM3}) and (\ref{RMM4}). In FFE, ${\bi E}$ is found from Maxwell's equations with $\rho{\bi E}+{\bi J}\times{\bi B}=0$ and $E_\parallel=0$, and for assumed boundary conditions. Such solutions are derived without consideration as to whether the plasma can provide the required $\rho$ and  ${\bi J}$.

The relevant response of the plasma to this background electric field corresponds to the low-frequency, long-wavelength limit. Although the self-consistent ${\bi E}$ is not known in general, one can write down the response to it. The response is  quite different for the perpendicular and parallel components. It is convenient to separate into these two components by writing
\be
{\bi E}=-{\bi u}\times{\bi B}+E_\parallel{\bi b},
\qquad
{\bi u}={{\bi E}\times{\bi B}\over B^2}.
\label{utimesB}
\ee
The response of individual particles to ${\bi E}_\perp=-{\bi u}\times{\bi B}$ may be described in terms of orbit theory, and the collective response to ${\bi E}_\perp$ may be approximated by the low-frequency long-wavelength limit of the plasma response tensor. 

The response of the plasma to $E_{\rm ind\parallel}$ is strongly oscillatory \citep{Letal05}, and there is a strong tendency to set up a potential field whose parallel component is equal and opposite to $E_{\rm ind\parallel}$, so that the self-consistent field has $E_\parallel\approx0$. Before discussing how such screening of $E_{\rm ind\parallel}$ occurs, let us consider the response in a region with $E_\parallel=0$.

\subsection{Drift motions}

In a region with $E_\parallel=0$, individual particles in the plasma respond to ${\bi E}$ through drift motions. The  electric drift velocity, ${\bi v}_E={\bi u}$ with ${\bi u}$ given by (\ref{utimesB}), applies to all particles. The electric drift is derived as the first-order term in an expansion of Newton's equation of motion for a charge in crossed electric and magnetic fields. With $E_\parallel=0$ the perpendicular equation of motion,
\be
m\gamma{d{\bi v}_\perp\over dt}=q({\bi E}_\perp+{\bi v}_\perp\times{\bi B}),
\label{drift1}
\ee
is approximated by expanding ${\bi v}_\perp$ in powers of $1/B$. The zeroth-order term is zero in a pulsar magnetosphere, due to the perpendicular momentum being radiated away in the superstrong magnetic field. Writing the first- and second-order terms as ${\bi v}_\perp={\bi u}+{\bi v}_{\rm pol}$, the first order solution of (\ref{drift1}) gives ${\bi u}={\bi v}_E$. The second-order term gives the polarization drift
\be
m\gamma{d{\bi u}\over dt}=q{\bi v}_{\rm pol}\times{\bi B},
\qquad
{\bi v}_{\rm pol}=-{m\gamma\over qB^2}\,{d{\bi u}\over dt}\times{\bi B}.
\label{drift2}
\ee
The polarization current density is given by integrating $q{\bi v}_{\rm pol}$ over the distribution of particles and summing over all species. 

Two other familiar drift motions, the grad-$B$ and grad-$P$ drifts, are $\propto v_\perp^2$, and hence are identically zero in a pulsar plasma. The curvature drift is non-zero, and its effect is to cause particles (with $v_\perp=0$) to follow the curved field lines \citep{SCE75}. Only the electric and polarization drifts are relevant here.

\subsection{Cold-plasma response}

In the low-frequency long-wavelength limit, the response may be described by that of a cold plasma, with appropriate reinterpretations of the cyclotron and plasma frequencies to include relativistic effects.  The induced current density has components
\bea
&&{\bi J}_{\rm pol}={c^2\over v_A^2}\varepsilon_0{\partial{\bi E}_\perp\over\partial t},
\qquad
{\bi J}_{\rm H}=\rho\,{{\bi E}_\perp\times{\bi B}\over B^2},
\nn\\\ms
&&\qquad
{\partial J_\parallel\over\partial t}=\varepsilon_0\omega_p^2E_\parallel+\sigma_\parallel{\partial E_\parallel\over\partial t}.
\label{pfc1a}
\eea
The first of the currents (\ref{pfc1a}) reproduces the polarization current derived from the polarization drift (\ref{drift2}) for a constant magnetic field, with the Alfv\'en speed in a pulsar magnetosphere satisfying $v_A^2\gg c^2$. The second of the currents (\ref{pfc1a}) is the Hall current, with the obvious interpretation of the charge density drifting at velocity ${\bi u}$, cf.\ (\ref{utimesB}). 

The parallel response is oscillatory, with $J_\parallel$ and $E_\parallel$ tending to oscillate at the plasma frequency, $\omega_p$.  Dissipation associated with the parallel response is included in (\ref{pfc1a}) through a parallel conductivity, $\sigma_\parallel$. In a cold plasma dissipation is associated with an electron collision frequency, $\nu_c$, which provides a frictional drag on the electrons, and implies a conductivity $\sigma_\parallel=\varepsilon_0\omega_p^2/\nu_c$. The generalization of the parallel response to a pulsar plasma retains these two features: oscillation and dissipation, with $J_\parallel$ and $E_\parallel$ oscillating at a relativistically modified plasma frequency. The relevant dissipation is collisionless and due to acceleration of particles, and this requires a detailed model. It is conventional to simulate the effect of acceleration through an anomalous conductivity, defined by replacing $\nu_c$ by an effective collision frequency, $\nu_{\rm eff}$.

\subsection{Screening of $E_\parallel$}

It is impossible for charges to screen an inductive electric field, but charges can flow freely along field lines to screen the parallel component, $E_{\rm ind\parallel}$. The effectiveness of such screening can be estimated by considering the parallel component of the wave equation (\ref{e1}). Assuming the parallel response (\ref{pfc1a}) and ignoring the conductivity, this becomes
\be
\left({1\over c^2}{\partial^2\over\partial t^2}-{\omega_p^2\over c^2}-\nabla_\perp^2-{\partial^2\over\partial s^2}\right)E_\parallel=-{1\over\varepsilon_0}{\partial\rho\over\partial s}
-\mu_0{\partial{\bi J}_{\rm ext\parallel}\over\partial t},
\label{e1p}
\ee
where $s$ denotes distance along the field line, and where ${\bi J}_{\rm ext}$ is any extraneous current not included in the response (\ref{pfc1a}). For the slowly (spatially and temporally) varying part of $E_\parallel$, the term involving $\omega_p^2/c^2$ dominates on the left-hand side. Supposing that $E_{\rm ind\parallel}$ varies over a characteristic distance $\ell_\parallel$, the right-hand side is of order $E_{\rm ind\parallel}/\ell_\parallel^2$. It follows that the screened $E_\parallel$ is smaller than  (the unscreened) $E_{\rm ind\parallel}$ by a factor of order  $c^2/\omega_p^2\ell_\parallel^2\ll1$, implying that screening of $E_{\rm ind\parallel}$ is very effective.

A localized region with $E_\parallel\ne0$ may be attributed to anomalous conductivity, with
\be
E_\parallel={J_\parallel\over\sigma_\parallel}={c^2\nu_{\rm eff}\over\omega_p^2}\,\mu_0J_\parallel,
\label{r1}
\ee
and with the anomalously conductivity identified as  $\sigma_\parallel=\varepsilon_0\omega_p^2/\nu_{\rm eff}$. This leads to a power dissipated per unit volume is $E_\parallel J_\parallel=J_\parallel^2/\sigma_\parallel$, which needs to be equated to the power per unit volume transferred to particles due to the acceleration of charges by $E_\parallel$. Although such a model based on anomalous conductivity is simplistic, two general conclusion based on it are likely to be valid more generally: gaps with $E_\parallel\ne0$ are highly localized, and effective dissipation within gaps is associated with acceleration of particles by $E_\parallel$.

\section{Role of gaps}
\label{sect:gaps}

An essential ingredient in any global model for pulsar electrodynamics is the presence of regions with $E_\parallel\ne0$. Here we give a general argument as to why $E_\parallel=0$ cannot apply everywhere, and then discuss specific gap models for aligned and oblique rotators.

\subsection{Need for gaps}
\label{sect:need}

One cannot have $E_\parallel=0$ everywhere in an obliquely rotating pulsar magnetosphere because this would lead to an inconsistency with the integrated form of Faraday's equation in the presence of the time-changing magnetic field. The integral of ${\bi E}$ along any closed path is equal to minus the rate of change of the enclosed magnetic flux. One may separate any closed path into a sets of closed subpaths, each of which includes paths along two field lines, and two paths across the field lines joining these two field lines. If one assumed $E_\parallel=0$ then the field lines are equipotentials and this integral is trivially zero around any such subpath. A time-changing magnetic field implies that that there must be some regions with $E_\parallel\ne0$ and that the field lines that pass through such regions are not equipotentials. The regions with $E_\parallel\ne0$ are identified as gaps.

The familiar frozen-in condition does not apply within a gap. The plasma motion above a gap may be regarded as slipping across field lines relative to the plasma motion below the gap. Such slippage is driven by the stress of the wind dragging the plasma in the inner magnetosphere backwards relative to corotation. This stress must be electrodynamic, cf.\ (\ref{Ms1}), (\ref{Ms3}), and communicated by parallel currents between the wind and the stellar surface \citep{S91}. However, such arguments do not provide any useful constraint on the possible locations of gaps \citep{BS10,YS12}.

\subsection{Charge starvation and gaps}
\label{sect:starvation}

The concept of charge starvation arose in models for an aligned rotator in which the only source of charge is the stellar surface. If the plasma is incapable of meeting the requirement on $\rho$ everywhere along a field line, it is assumed that a region with $E_\parallel\ne0$ develops and results in pair creation, providing the necessary additional source of charges. 

A `vacuum gap' with $E_\parallel\ne0$, first proposed by \cite{RS75}, became the basis for `inner' gap models. The concept of a gap was later generalized to include a slot gap \citep{AS79} and an outer gap \citep{CHR86}. \cite{RS75} assumed that the plasma above the vacuum gap is rotating at a lower angular speed, $\omega^*=\omega-\Delta\omega$ say, to the corotating plasma below the gap. In the aligned case, the corotation potential (\ref{RMM2}) becomes $\Phi_{\rm cor}=\pm\mu_0m\omega\sin^2\theta/4\pi r=\pm\mu_0m\omega/4\pi r_0$, where $r_0=r/\sin^2\theta$ is the field-line constant, implying that $\Phi_{\rm cor}$ is constant along a given field line (given $r_0$). In the presence of  a gap, there is a potential difference $\Delta\Phi=\mu_0m\Delta\omega/4\pi r_0$, along a field line through the gap. $\Delta\Phi$ is zero along the axis ($r_0\to\infty$, $\sin\theta=0$) and is maximum at the last closed field line  ($r_0=r_L$, $\sin^2\theta=r/r_L$), favouring a slot-gap model. An outer gap is needed to provide additional charges near the point where $\rho_{\rm cor}$ changes sign which, for an aligned rotator, is at $r=2r_0/3$ and $\cos^2\theta=1/3$. 

In brief, the corotation charge density cannot be provided by charges from the stellar surface alone, requiring a purely magnetospheric source of charge, which is attributed to pair creation by charges accelerated by $E_\parallel\ne0$. Further arguments suggest that pair creation in the magnetosphere is the dominant source of charge, and that charges from the stellar surface may play no significant role \citep{T10}.

\subsection{Gaps in an oblique rotator}

A generalization of the foregoing gap model to an oblique rotator is possible. Suppose that one has $E_\parallel=0$ between gaps. For a dipolar field, one can ensure $E_\parallel=0$ by assuming the electric field of the form
\be
{\bi E}={\bi E}_{\rm cor}-\grad\Phi'(\chi,\phi_0),
\label{ob1}
\ee
where $\chi=\sin^2\theta_b/r$, $\phi_0=\phi_b$ are the field-line constants for a particular dipolar field line, in terms of the polar and azimuthal angles relative to the magnetic axis. A model for a specific gap involves specifying the function $\Phi'(\chi,\phi_0)$ on either side of the gap. A complication, compared with a gap in an aligned model, is that the potential drop across the gap includes the contribution from the line integral of ${\bi E}_{\rm ind}$ through the gap, in addition to the change in $\Phi'(\chi,\phi_0)$ from below to above the gap.

\subsection{Drifting subpulses}
\label{sect:subpulses}

A potentially attractive feature of a gap model in an oblique rotator relates to drifting subpulses.  Observations of  drifting subpulses imply azimuthally dependent structures drifting at an angular velocity different from that of the star. In the carousel model \citep{DR99} drifting subpulses are associated with magnetospheric density structures $\propto\cos m\phi_b$ where $m$ is an integer. The possible description in terms of a gap in an oblique rotator is that $\Phi'$ can change from being independent of $\phi_b$ below the gap to $\propto\cos m\phi_b$ above the gap. A specific plasma instability within the gap is needed in any detailed model for the development of such azimuthal structures, and several have been suggested \citep{KMM91}. For example, a gradient in the flow velocity, as a function of $\theta_b$, implies a shear that can lead to a diocotron instability \citep{PHB02}, resulting in the development of an azimuthally asymmetric structure \citep{S04,FKK06}. This instability could lead to a structure dominated by a particular $Y_{lm}(\theta_b,\phi_b)$ with a large $l,m$, as suggested by the carousel model.

\subsection{Large-amplitude longitudinal oscillations (LALOs)}
\label{sect:LALOs}

Static models for gaps are unstable to temporal perturbations  \citep{Letal05}. This does not invalidate the requirement for regions with $E_\parallel\ne0$, but it does require a reinterpretation of what a gap is. 

The parallel response of a pulsar plasma, consisting of streaming, relativistic electrons and positrons with $p_\perp=0$, differs in detail from that of a cold plasma, but retains the important feature that it is oscillatory at the generalization of $\omega_p$, which includes a factor $\langle\gamma\rangle^{1/2}$ in the denominator, where $\langle\gamma\rangle$ is a weighted average of the Lorentz factor of the particles. Specific models show that large-amplitude longitudinal oscillations (LALOs) develop \citep{Letal05,BT07} with a saw-tooth profile for $E_\parallel$. One suggestion is that `gaps', in the sense of regions with $E_\parallel\ne0$ where effective acceleration occurs, be regarded as propagating LALOs \citep{LM08} rather than quasi-stationary structures. 

\subsection{Current starvation}
\label{sect:current_starvation}

Any model for pulsar electrodynamics can be valid only if the plasma is able to supply not only the required charge density, but also the required current density. `Current starvation' refers to situations where the plasma cannot supply the current density ${\bi J}_{\rm cor}$, given by (\ref{RMM4}). The current ${\bi J}_{\rm cor}$ is required to maintain the time-changing $\rho_{\rm cor}$ at its instantaneous value. The current density (\ref{pfc1a}) that the plasma can provide is different in the three orthogonal directions, ${\bi E}_\perp, {\bi E}\times{\bi B}, {\bi B}$. Ignoring angular factors, one finds that ${\bi J}_{\rm cor}$ is of order $\rho_{\rm cor}\omega r$, which is the same order as the response along ${\bi E}\times{\bi B}$. The polarization current, which is the response along  ${\bi E}_\perp$, includes the very small factor $c^2/v_A^2\ll1$. As a consequence, the plasma cannot provide the required current along ${\bi E}_\perp$. Specifically, ${\bi J}_{\rm cor}$ is determined by (minus) the displacement current, due to ${\bi E}_{\rm pot}={\bi E}_{\rm cor}-{\bi E}_{\rm ind}$, whereas ${\bi J}_{\rm pol}$ is equal to $c^2/v_A^2$ times ${\bi J}_{\rm disp}$. It follows that $c^2/v_A^2\ll1$ implies $|{\bi J}_{\rm pol}|\ll|{\bi J}_{\rm cor}|$. This would appear to invalidate the RMM in the oblique case. However, there is another way in which the time-changing $\rho_{\rm cor}$ could be maintained. 

The required cross-field current could be supplied by a current flow along field lines to a conducting surface, here the stellar surface, closure across field lines at the surface, and return current flow along neighboring field lines. This form of cross-field current flow has been invoked in connection with laboratory plasmas \citep{S55}, and with the `current wedge' in a (terrestrial) substorm \citep{McP78}. However, such current flow can be effective in a pulsar only near the surface of the star. At distances that are a significant fraction of $r_L$, the lapse time associated with propagation (at $c$) to the surface of the star and back becomes comparable with the pulsar period, $P$, which is also the time scale on which $\rho_{\rm cor}$ is required to change. This way of providing the required ${\bi J}_{\rm cor}$ is possible only if the lapse time $\approx2(r-R_*)P/r_L$ is much smaller than $P$. When this condition is not satisfied, there is current starvation, and $(\bomega\times{\bi x})_\perp$ cannot be maintained at its required instantaneous value. Corotation, even in the restricted sense implied by ${\bi v}_{\rm cor\perp}$, cf.\ (\ref{RMM1}), is then not possible and the RMM is invalid. 

\subsection{Partial current starvation}

A simple model for the case where there is partial current starvation involves first considering the case where the time-varying part of $\rho_{\rm cor}$ is absent. This case is given by replacing  $\rho_{\rm cor}$ and ${\bi E}_{\rm pot}={\bi E}_{\rm cor}-{\bi E}_{\rm ind}$ by their time averages $\langle\rho_{\rm cor}\rangle$ and $\langle{\bi E}_{\rm pot}\rangle$, respectively; this is achieved by the replacement ${\hat{\bi m}}\to{\hat{\bi m}}-{\hat{\bi m}}\cdot{\hat{\bomega}}\,{\hat{\bomega}}$ in (\ref{RMM3}) and (\ref{Epot}), respectively. Introducing a parameter $0\le y'\le1$ to describe the degree of current starvation, the model corresponds to replacing $\rho_{\rm cor}$ and ${\bi E}_{\rm pot}$ by
\be
\rho'_{\rm cor}=(1-y')\rho_{\rm cor}+y'\langle\rho_{\rm cor}\rangle,
\qquad
\qquad
{\bi E}'_{\rm pot}=(1-y'){\bi E}_{\rm pot}+y'\langle{\bi E}_{\rm pot}\rangle,
\label{cs1}
\ee
respectively.  The parallel component of the inductive electric field, $E_{\rm ind\parallel}$, is screened by $E_{\rm pot\parallel}$ for $y'=0$, and this screening is incomplete for $y'\ne0$. There is a non-zero parallel electric field in this model:
\be
E'_\parallel=y'(E_{\rm ind\parallel}+\langle E_{\rm pot\parallel}\rangle).
\label{cs2}
\ee

The foregoing arguments imply that the parameter $y'$ is of order $r/r_L$. It follows that current starvation increases in significance as $r/r_L$ increases. The retarded terms in the VDM, which are neglected in the discussion here are of order $r/r_L$, and in any detailed model for current starvation, the retarded terms need to be taken into account. An interesting feature of a gap due to this current starvation is that, unlike gaps that form due to charge starvation, such a gap is not restricted to the open-field region.

\subsection{Location of source regions}

A reliable model for the location of gaps is required to make predictions concerning the acceleration of particles and the resulting pair creation. Emphasis has been placed on the location of gaps that lead to the emission of high-energy photons. Even for the high-energy emission, the electrodynamics does not lead to a convincing identification of the location of the gaps \citep{BS10}, but it is plausible that the acceleration and emission regions are colocated.  Identifying the location of the radio source region is more problematic because the relationship between the acceleration region and the radio emission region is not known. For the two established coherent emission mechanisms, cf. \S\ref{sect:emission}, this relationship is (observationally) quite different. The acceleration and emission regions are different but closely correlated for electron cyclotron maser emission, and are not only different but can be very widely separated for plasma emission. Attempting to identify the source region for pulsar radio emission from first principles seems unrealistic.

\section{Wave dispersion in pulsar plasma}
\label{sect:dispersion}

Wave dispersion in any magnetized plasma is determined by the plasma response tensor, which depends on the distributions of particles. 

\subsection{Pulsar plasma}

Unusual features of a pulsar plasma, in the polar-cap regions, include: the plasma is dominated by relativistic electrons and positrons that are streaming outward with a mean Lorentz factor $\langle\gamma\rangle\gg1$; there is also a relativistic spread in Lorentz factor about this mean; the plasma in one-dimensional (1-D), in the sense that all particles are in their ground Landau state, corresponding classically to $p_\perp=0$; the cyclotron frequency, $\Omega_\rme$, is much higher than the plasma frequency. 

In earlier literature it was assumed that `primary' particles are accelerated from the stellar surface to very high Lorentz factors, $10^6$--$10^7$, and that these trigger the pair creation of the much more numerous `secondary' pairs. The number densities, $n_\pm$, of the secondary positrons and electrons, respectively, satisfy $e(n_+-n_-)=\rho$, and a multiplicity, $M$, may be defined by writing $\half(n_++n_-)=M|\rho|/e$, with $\rho=\rho_{\rm cor}$ in the RMM.  The distributions of electrons and positrons have been estimated from numerical models of the pair creation \citep{HA01,AE02}. The numbers have considerable uncertainty, due to uncertainties in the parameters assumed in the numerical models. Estimates suggest $M$ in the range of a few tens to $10^3$, and streaming Lorentz factors in the range $\langle\gamma\rangle=10^2$--$10^3$ with a thermal like spread about this mean \citep{AE02}. 

\subsection{Cold, relativistically streaming pair plasma}
\label{sect:cold}

A simple model for wave dispersion in a pulsar plasma involves assuming that, in the rest frame of the plasma, the electrons and positrons are cold \citep{MS77,AB86,BA86}. Wave dispersion in this frame may be regarded as a modification of the magnetoionic theory, which applies to a cold electron gas, to include a mixture of electrons and positrons. The wave properties in the pulsar frame may be found by solving for the wave properties in the rest frame and Lorentz transforming these properties to the pulsar frame. 

The mixture of electrons and positrons may be described by a parameter $\epsilon$, which is the average charge per particle, with $\epsilon=-1$ for a pure electron gas, $\epsilon=0$ for a pure pair plasma. Dispersion in an electron gas, $\epsilon=-1$, is conventionally described in terms of two magnetoionic parameters, denoted $X=\omega_p^2/\omega^2$, $Y=\Omega_\rme/\omega$ here, with $\epsilon$ being a third parameter in the case of a mixture. The dispersion equation becomes a quadratic equation for $n^2$ (the square of the refractive index) as a function of the angle, $\theta$, between the wave normal and the magnetic field, in addition to $X,Y,\epsilon$. In a pulsar plasma, $\Omega_\rme$ is much higher than all other frequencies of interest and one is justified in expanding in inverse powers of $\Omega_\rme$, which is equivalent to expanding in $X/Y^2=\omega_p ^2/\Omega_\rme^2=1/\beta_A^2$, where $\beta_A=v_A/c$. One has $\beta_A^2\gg1$ in a pulsar plasma, and then the characteristic speed of MHD waves is $\beta_Ac/(1+\beta_A^2)^{1/2}\approx c$.

The dispersion equation for a cold pair plasma with $\epsilon\ne-1$ is, to first order in $1/\beta_A^2$,
\bea
&&(n^2-1)[(1-X\cos^2\theta)n^2-(1-X)]+{1\over \beta_A^2}\bigg\{
n^4\sin^2\theta
-[(2-\epsilon^2X)\sin^2\theta
\nn
\\
&&\qquad\qquad
+(1-X)(1+\cos^2\theta)]n^2
+(1-X)(2-\epsilon^2X)\bigg\}=0.
\qquad
\label{infB1}
\eea
To zeroth order in $1/\beta_A^2$  for the ordinary (o) mode and to first order for the extraordinary (x) mode, (\ref{infB1}) gives
\be
n_{\rm o}^2={1-X\over1-X\cos^2\theta}
={\omega^2-\omega_p ^2\over\omega^2-\omega_p ^2\cos^2\theta},
\qquad
n_{\rm x }^2=1+{1\over\beta_A^2}.
\label{infB2}
\ee
The o~mode separates into a low-frequency branch, $\omega^2<\omega_p ^2\cos^2\theta$, and a high-frequency branch $\omega^2>\omega_p ^2$. There is a resonance at $\omega^2=\omega_p ^2\cos^2\theta$ in the lower branch and a cutoff at $\omega^2=\omega_p ^2$ in the upper branch,  which are separated by a stop band, $\omega_p ^2\cos^2\theta<\omega^2<\omega_p ^2$. For the o~mode, at sufficiently low frequencies, the dispersion relation approaches $\omega=|k_z|c$, which corresponds to an Alfv\'en wave in the strong-field limit $\beta_A\to\infty$. At high frequencies, $\omega_p\ll\omega\ll\Omega_\rme$, the dispersion relation approaches $n_{\rm o}^2\approx1-(\omega_p ^2/\omega^2)\sin^2\theta$, which is equivalent to $\omega^2-|{\bi k}|^2c^2\approx\omega_p ^2\sin^2\theta$. The polarization of the o~mode is mixed longitudinal transverse, with the transverse component along the projection of the wave vector across the magnetic field. The x~mode is vacuum-like, $\omega=kc$, to zeroth order in $1/\beta_A^2$; its polarization is transverse, perpendicular to both the wave vector and the magnetic field. 

On including the first-order term in $1/\beta_A^2$ in (\ref{infB1}), the two modes become elliptically polarized. In the limit $\omega\gg\omega_p $ the transverse polarization is described by its axial ratio, $T$, with $T=T_\pm$ for the two modes satisfying
\be
T_\pm=\half R
\pm 
\half(R^2+4)^{1/2},
\qquad
R={\Omega_\rme\sin^2\theta\over\epsilon\omega\cos\theta}.
\label{SPP8}
\ee
The modes are nearly circularly polarized, $|T_\pm|\approx1$, for $Y^2\sin^4\theta\ll4\epsilon^2\cos^2\theta$, and nearly linearly polarized for $Y^2\sin^4\theta\gg4\epsilon^2\cos^2\theta$. 

The observed polarization of single pulses \citep{MS00,J04,ES04} is strongly indicative of propagation through a birefringent medium with elliptically polarized modes \citep{Metal06}. In principle one could use the ellipticity of the observed polarization to estimate $R$, and hence to constrain the plasma parameters through a requirement on $R$, cf.\ (\ref{SPP8}). In practice, this approach has not led to useful constraints.

\subsection{Lorentz transformation to the pulsar frame}

The foregoing wave properties apply to a cold pair plasma in its rest frame, and a model for dispersion in a pulsar plasma is obtained by applying a Lorentz transformation to the pulsar frame, in which the plasma is streaming relativistically with $\gamma\gg1$. The dispersion may be treated either by solving for the wave properties in the and transforming them to the pulsar frame, or by transforming the response tensor  \citep{M13} to this frame and solving for the wave properties in the the pulsar frame. Here we ignore the wave dispersion, in which case the Lorentz transformation implies the formulae for the aberration of light. With quantities in the pulsar frame be indicated by a prime, this gives
\be
\omega=\gamma\omega'(1-\beta\cos\theta')\approx{\omega'\over2\gamma}(1+\gamma^2\theta'^2),
\qquad
\cos\theta={\cos\theta'-\beta\over1-\beta\cos\theta'}\approx{\gamma^2\theta'^2\over1+\gamma^2\theta'^2},
\label{LT1}
\ee
where the approximate forms apply for $\theta'^2\ll1$, $\gamma^2\ll1$.

The Lorentz transformation (\ref{LT1}) has the following semi-quantitative effects. The forward hemisphere, $\theta<\pi/2$, in the rest frame transforms into a forward cone $\theta'\lesssim1/\gamma$ in the primed frame. The escaping radiation with $1/\gamma\lesssim\theta'<\pi/2$ arises from $\pi/2<\theta\lesssim\pi-1/\gamma$, so that most of the escaping radiation corresponds to backward-propagating waves in the rest frame. The ratio of the observational frequency, $\omega'$, to the frequency of the radiation in the rest frame is large, $\omega'/\omega\approx\gamma$, for most angles $\theta'\gg1/\gamma$, and is small, $\omega'/\omega\lesssim1/\gamma$, only for $\theta'\ll1/\gamma$. 

In summary,  if the plasma in the emission region is streaming outward with $\gamma\gg1$, as is usually assumed, then the observed emission arises from emission in the rest frame predominantly in the backward direction at a frequency of order $1/\gamma$ times the observed frequency. The amount of Lorentz boosting of the frequency adds another uncertainty in relating the observed frequency to the intrinsic frequency of proposed emission mechanisms.

\subsection{Dispersion in intrinsically relativistic pulsar plasmas}

In a thermal plasma, the cold-plasma approximation is valid when the phase speed of the waves is much greater  than the mean (thermal) speed of particles. In a pulsar plasma, the mean speed of the particles is close to $c$. A relevant counterpart of the mean speed is the weighted mean $\delta\beta^2=\<\gamma \beta^2\>/\<\gamma\>$, where the angular brackets denote the average over the 1D distribution function.

An important special case when considering the radio emission is the generalization of Langmuir waves in a thermal plasma. The dispersion relation for  parallel-propagating longitudinal waves, denoted as the $L$-mode, is \citep{Metal99}
\be
\omega=\omega_L(z),
\qquad
\omega_L^2(z)=\omega_p ^2z^2W(z),
\label{7f.19}
\ee
with 
\be
W(z)=\left\<{1\over\gamma^{3}(z-\beta)^{2}}\right\>,
\qquad
z={\omega\over k_zc},
\label{Wz}
\ee
when the spread in parallel momenta is taken into account. The $L$~mode has a cutoff at $\omega=\omega_{\rm c}$ and crosses the light line at $\omega=\omega_1$:
\be
\omega_{\rm c}^2=\omega_L^2(\infty)=
\omega_p ^2\left\<\gamma^{-3}\right\>,
\qquad
\omega_1^2=\omega_L^2(1)=
\omega_p ^2\left\<\gamma\right\>
(1+\delta\beta^2).
\label{7f.20}
\ee
As for Langmuir waves in a nonrelativistic plasma, the parallel $L$~mode has a maximum frequency at a phase speed of order the mean speed of the particles, $\sim(\delta\beta^2)^{1/2}c$, which is very close to the speed of light in a highly relativistic plasma. Landau damping results from resonance at $\omega=k_zv$, and is strong for phase speeds near and below the mean speed of the particles. As a consequence, the $L$~mode effectively ceases to exist for phase speeds near and below this maximum. 

An unusual feature of the $L$-o~mode is that it can be generated through Cerenkov emission, and hence through a streaming instability, as for Langmuir waves generated by an electron beam propagating through a thermal plasma.  (Cerenkov emission of the x~mode is not possible, due to the polarization of the x~mode having zero component along the magnetic field.) A dispersion curve the $L$-o~mode can cross the light line twice, so that its phase speed becomes subluminal over a range of parameters. That is, the combination of anisotropy and highly relativistic particles in a pulsar plasma allows the $L$-o~mode to change from superluminal to subluminal and back to superluminal along a range of dispersion curves \citep{Metal99}. Subluminal waves can be generated by a beam instability. For a dispersion curve that joins on continuously to the (superluminal) high-frequency o~mode, these waves can escape \citep{MG99,Metal99}. Such direct escape has no counterpart for Langmuir waves in a thermal plasma, due to the dispersion equation separating into separate equations for longitudinal and transverse waves in an isotropic plasma; longitudinal waves can produce escaping transverse radiation only through scattering or mode coupling. It also has no counterpart in a cold electron gas, where the only waves that can escape directly are in the o~and x~modes, which have refractive indices less than unity. This unusual feature of the dispersion of the $L$-o~mode is the basis for one of the possible pulsar radio emission mechanisms discussed in \S\ref{sect:emission}.

\section{Radio emission mechanisms}
\label{sect:emission}

Pulsar radio emission in one of three recognized coherent emission processes in astrophysical plasmas, with the other two being plasma emission and electron cyclotron emission (ECME). Unlike the other two, there is no widely accepted theory for the pulsar radio emission mechanism. The two most favoured mechanisms are probably coherent curvature emission and some form of plasma emission. Two others are linear acceleration emission and anomalous Doppler emission. It is possible in principle that there may be more than one effective emission mechanism operating simultaneously, but this seems unlikely because it would require two separate mechanisms to produce emission at similar frequencies with similar extreme brightness. In the following discussion it is assumed that there is only one pulsar radio emission mechanism.

\subsection{Classification of radio emission mechanisms}

In an early review of pulsar radio emission mechanisms, \cite{GZ75} emphasized the effective brightness temperature, $T_{\rm eff}$, as a measure of coherence. The brightness temperature is a constant along the ray path from the source to the observer. To estimate it from observation requires an estimate of the area, perpendicular to the line of sight, from which the radiation is emitted in the source. Subject to the large uncertainties involved, $T_{\rm eff}$ for sources of plasma emission and sources of ECME are typically between $10^{10}\,$K and $10^{20}\,$K, whereas pulsar emission is much brighter, $10^{25}\,$K to $10^{30}\,$K. In extreme cases of very short pulses, of duration $\delta t$ say, an argument that limits the area of emission is that it cannot have a linear dimension greater than the light-propagation time $c\delta t$. For nanosecond structures observed in radio emission from the Crab pulsar \citep{HE07}, this implies a source size of order $1\,$m, and $T_{\rm eff}>10^{40}\,$K. Less extreme examples, such as micropulses, suggests that pulsar emission consists of many extremely bright, transient, localized bursts of emission. A detailed theory needs to be capable of accounting for such extremely bright fine structures in the emission, as well as providing a semi-quantitative basis for the average emission, presumably consisting of a statistically large number of such fine structures.

\cite{GZ75} discussed two classes of `coherent' emission mechanisms: antenna and maser mechanisms. In the simplest form of an antenna mechanism, a bunch of $N$ particles in a sufficiently small volume radiates $N^2$ times the power per particle. Antenna mechanisms may be further classified according to how the bunch is formed. In one class, which we refer to as emission by bunches, the existence of bunches is simply postulated: how the bunch is formed is not specified, and its dimensions are assumed small compared with a wavelength of the emitted radiation. The other class involves growth of the wave due to self-bunching. A maser mechanism involves negative absorption. These three classes of coherent emission may be distinguished by the form of the distribution of particles. For the first class, involving prebunched particles, the postulated bunch is confined to small regions in both coordinate space and momentum space. For a self-bunching instability, the initial localization is only in momentum space. A maser instability is driven by some form of inverted energy populations rather than by bunching. The back reaction to the emission in each case tends to reduce the feature that drives the coherent emission. For example, the back reaction to coherent emission by a bunch tends to disperse the bunch  in coordinate space, so that the instability suppresses itself. 

Self-bunching mechanisms and maser mechanisms may be interpreted as reactive and resistive plasma instabilities, attributed to the reactive and resistive parts, respectively, of the plasma response tensor. In the case of plasma emission, the instability involves growth of Langmuir waves in a beam instability. If the initial spread in momentum space is sufficiently small, a reactive versions of the instability develops, associated with self-bunching along the beam axis.  Two different forms of self-bunching are  possible in cyclotron instabilities \citep{W83}, but only axial bunching is possible in a 1-D pulsar plasma. The back reaction to the self-bunching instability broadens the spread in momentum space, until the instability suppresses itself. As a result the self-bunching instability evolves into a maser instability. 

A maser theory applies when the random phase approximation (RPA) is valid, and this requires that the growth rate of the instability be less than the bandwidth of the growing waves. (A reactive instability applies when this inequality is reversed.) In models for plasma emission in solar radio bursts and for sources of ECME, the growth rates appear consistent with the requirements for the RPA to apply, and antenna mechanism are not relevant. However, a maser version of coherent emission seems inconsistent with the brightest fine structures, such as nanopulses,  in pulsar emission. Very bright fine structures required large growth rates, seemingly inconsistent with the RPA.

A complementary approach to interpreting any specific form of coherent emission concerns the free energy that drives the relevant plasma instability. Plasma emission involves a beam instability and the free energy is associated with the beam of fast electrons moving through the background plasma. The maser version is the bump-in-tail instability, driven by a distribution with $\partial f/\partial v_\parallel>0$, where $v_\parallel$ is the component along the direction of the beam. ECME is driven by a distribution with $\partial f/\partial v_\perp>0$, and this is not relevant in a pulsar magnetosphere due to $v_\perp=0$. There are two sources of free energy for a distribution with $p_\perp=0$. One is a distribution with $\partial f/\partial\gamma>0$, with $\gamma\approx p_\parallel/mc$, which is a relativistic counterpart of a bump-in-tail distribution. The other source of free energy is the extreme form of anisotropy, $f\propto\delta(p_\perp)$, implied by the 1-D distribution.

\subsection{Coherent curvature emission}
\label{sect:curv}

Curvature emission by relativistic particles is similar to synchrotron emission, in the sense that both may be modelled in terms of a relativistic particle moving along the arc of a circle. In curvature emission by particles with $p_\perp=0$, the radius of the circle corresponds to the radius of curvature, $R_c$, of the magnetic field line. The frequency of curvature emission from an individual particle has a maximum around $\omega_c$ of order $c\gamma^3/R_c$. Curvature emission was invoked in two different ways in early pulsar models. One purpose is as a high-frequency emission mechanism, whereby the primary particles emit gamma rays that decay into pairs. For this purpose, a dipolar field line near the polar caps has a radius of curvature that is too large for $\hbar\omega_c$ to plausibly account for the gamma-ray energy required, and it was argued that a much smaller $R_c$ requires that the magnetic field has substantial multipole components. The other purpose is as a radio emission mechanism, which applies at low frequencies, $\omega\ll\omega_c$, where the single-particle power spectrum for curvature emission increases $\propto\omega^{1/3}$. The coherence was postulated to result from a (reactive) beam instability leading to self-bunching \citep{RS75}. 

An argument against the antenna mechanism for curvature emission concerns the generalization of the idealized case of $N$ particles emitting $N^2$ times the power per particle to a  more realistic bunch. Consider a bunch of $N$ particles all with the same velocity with a spatial distribution $n({\bi x})$ about their mean instantaneous position.  For any type of emission, the power in the range $d^3{\bi k}/(2\pi)^3$ about wave vector ${\bi k}$, is enhanced by a factor $|{\tilde n}({\bi k})|^2$, where ${\tilde n}({\bi k})$ is the spatial Fourier transform of $n({\bi x})$. The idealized case of a bunch with zero dimensions, corresponding to $n({\bi x})=N\delta^3({\bi x})$, implies $|{\tilde n}({\bi k})|^2=N^2$. For a more realistic model of a bunch, the power emitted in the range $d^3{\bi k}/(2\pi)^3$ is modified by the ${\bi k}$-dependence of the factor $|{\tilde n}({\bi k})|^2$.  For emission by highly relativistic particles, one has $k_\perp/k_\parallel$ of order $1/\gamma$, and a plausible model for a  bunch is a pancake with perpendicular to parallel dimensions of order $\gamma$ \citep{M92}. Moving along a curved field, such a pancake quickly ceases to be nearly perpendicular to the field line, and the peak in ${\bi k}$ in $|{\tilde n}({\bi k})|^2$ no longer coincides with the peak in ${\bi k}$ in the emission formula for a single particle, so that the coherence becomes ineffective. A different argument against coherent curvature emission by bunches was given by \cite{Letal98}. Despite such objections to it, coherent curvature emission continues to be applied to the interpretation of the emission from radio pulsars.

\subsection{Plasma emission mechanisms}
\label{sect:plasma_em}

Plasma emission, e.g., in type~III solar radio bursts, involves Langmuir waves being generated by a beam instability and then being partly converted into escaping transverse waves by nonlinear processes in the plasma. The brightness temperature of the resulting emission is restricted to less than the effective temperature of the Langmuir waves. Semi-quantitative models suggest that the nonlinear conversion processes saturating at this level can account for the brightness of type~III emission \citep{M86}. In contrast, to account for the very bright pulsar emission, a plasma-emission mechanisms requires both a growth rate larger than simple models suggest, and an efficient conversion mechanism.  The relatively inefficiency of conversion processes has been described as a `bottleneck' \citep{U00} for a pulsar plasma emission mechanism. Ways of overcoming both these problems have been suggested.

The beam instability invoked in earlier models involves a relatively low-density, very-high-energy beam of `primary' particles propagating through higher-density, lower-energy pair plasma, and the growth rate of this instability is too small to account for effective growth. One suggestion for overcoming this difficulty is to postulate that the pair creation is strongly non-stationary, producing `clouds' such that the faster particles in a following cloud can overtake the slower particles in a preceding cloud leading to the beam instability \citep{UU88}. Another suggestion is that the instability results in soliton-like `Langmuir microstructures' \citep{APS90,A93}, somewhat similar to the suggested LALOs discussed above in connection with screening of $E_\parallel\ne0$. 

The maximum growth rate for the beam instability occurs for a `resonant' form of the instability. The dispersion relation for longitudinal waves in a relativistic counterstreaming plasma reduces to a quartic equation for parallel propagation when the velocity spreads are neglected \citep{VM11}. Consider a beam with number density $n_b$ and velocity $\beta_bc$ propagating along ${\bi B}$ through a 1-D pair plasma with number density $n_e$. The dispersion equation for parallel longitudinal waves is given by (\ref{7f.19}), and for a cold beam one has $z^2W(z)=1+({n_b/ n_\rme}){z^2/\gamma_b^3(z-\beta_b)^2}$, so that (\ref{7f.19}) becomes
\be
1-{\omega_p ^2\over\omega^2}
-{n_b\over n_\rme}\,{\omega_p ^2\over\gamma_b^3(\omega-k_z\beta_b)^2}=0,
\label{8d.8a}
\ee
which is the quartic equation for $\omega$. For $n_b/\gamma_b^3\ll n_\rme$ the effect of the beam is significant only for $\omega\approx k_z\beta_b$. In the limit of arbitrarily large $k_z\beta_b$, the four solutions of (\ref{8d.8a}) approach $\omega=\pm\omega_p ,k_z\beta_b\pm\omega_p (n_b/n_\rme\gamma_b^3)^{1/2}$. The solution near $\omega=-\omega_p $ is of no interest, and it is removed by approximating the quartic equation (\ref{8d.8a}) by the cubic equation
\be
(\omega-\omega_p )(\omega-k_z\beta_b)^2
-{n_b\over2n_\rme\gamma_b^3}\,\omega_p \omega^2=0.
\label{8d.9}
\ee
The solutions of the cubic equation simplify in two forms of reactive instability: a `resonant' form $k_z\beta_b\approx\omega_p $, and a `non-resonant' form $\omega\ll\omega_p $. The approximate solutions for the growth rate in these two forms are \citep{GGM02a,GGM02b}
\be
\omega\approx
\left\{\begin{array}{ll}
{\displaystyle\omega_p 
+\rmi{\omega_p \over\gamma_b}{\sqrt{3}\over2}\left(\frac{n_b}{2n_\rme}\right)^{1/3}
 }, &\mbox{resonant},
\\
\ms
{\displaystyle k_z\beta_b
+\rmi{\omega_p \over\gamma_b^{3/2}}\left(\frac{n_b}{2n_\rme}\right)^{1/2} 
},
&\mbox{nonresonant}.
\end{array}
\right.
\label{8d.10}
\ee
Enhanced growth in the resonant case partly overcomes the problem of the growth rate being too low.

The problem of the `bottleneck' in the conversion process can be avoided entirely. A detailed investigation of the dispersive properties of the plasma \citep{MG99,Metal99} shows that the Langmuir-like mode for parallel propagation becomes the $L$-o~mode for oblique angles of propagation. This implies that the Langmuir-like mode joins on continuously to the escaping o~mode \citep{GGM02a,GGM02b}. It follows that no nonlinear conversion is needed to produce escaping o~mode radiation. Some variant of this form of plasma emission is perhaps the most plausible pulsar radio emission mechanism.

\subsection{Maser curvature emission}

A maser mechanism involves negative absorption. %A maser version of the beam instability is well known, but its growth rate is too slow to be relevant to pulsar radio emission. 
In the simplest approximation, the absorption coefficient corresponding to curvature emission is strictly positive \citep{B75,M78}. Although absorption can be negative when the curvature drift and field-line torsion are taken into account \citep{CS88,LM92,LM95}, the growth rate for such maser curvature emission seems too small for it to be an effective pulsar emission mechanism. Two other suggested maser emission mechanisms are linear acceleration emission and anomalous Doppler emission.

\subsection{Linear acceleration emission (LAE)}

Any accelerated motion of a charged particle leads to radiation, and acceleration by $E_\parallel\ne0$ leads to linear acceleration emission (LAE). In contrast with curvature emission, LAE involves acceleration along the magnetic field rather than perpendicular to the magnetic field. In the simplest case, where $E_\parallel$ is oscillating at a frequency $\omega_0$, LAE occurs at frequencies $\omega\lesssim\omega_0\gamma^2$. The emission by a charge with $p_\perp=0$ accelerated along a curved magnetic field line is curvature-like at $\omega\lesssim(c/R_c)\gamma^3$, and LAE-like for $(c/R_c)\gamma^3\lesssim\omega\lesssim\omega_0\gamma^2$ \citep{M78}.  

A maser form of LAE is a possible pulsar radio emission mechanism \citep{M78}. A realistic model for maser LAE needs to be based on a realistic model for the regions/structures with $E_\parallel\ne0$. It seems plausible that the regions with $E_\parallel\ne0$ can be modeled in terms of large-amplitude longitudinal waves \citep{APS90,A93,Letal05,BT07,LM08}, and some progress has been made in investigating LAE in such cases \citep{R92,R95,MRL09,ML09,RK10}. However, it has yet to be shown that maser LAE is a viable radio emission mechanism for pulsars.

\subsection{Anomalous Doppler instability}

Negative gyromagnetic absorption is the accepted mechanism for ECME, but it requires a distribution with $\partial f/\partial p_\perp>0$ , and this is not possible in a pulsar plasma due to all particles having $p_\perp=0$. Alternative gyromagnetic instabilities have been suggested for pulsars.

The gyroresonance condition may be written in the form \citep{KMM91}
\be
\gamma(\omega-k_\parallel v_\parallel-k_x u_{\rm drift})-s\Omega_\rme=0,
\label{res}
\ee
where the drift velocity, $u_{\rm drift}$,  is assumed to be along the $x$ axis. In a pulsar plasma, gyro\-magnetic absorption can be negative only for $s=0$, which corresponds to the Cerenkov resonance, and for $s=-1$, which is the anomalous Doppler resonance.  The resonances at both $s=0$ and $s=-1$ require $v_\parallel\ge\omega/k_\parallel$, and this is possible only for waves with phase speed, $\omega/k_\parallel$, less than $c$. Negative absorption at $s=0$ includes a maser version of the beam instability, cf. \S\ref{sect:plasma_em}. Two other maser instabilities based on (\ref{res}) were suggested in connection with pulsars by \cite{MU79}: a drift instability, which is not discussed here, and an anomalous cyclotron instability \citep{LBM99}. 

One may interpret the gyroresonance condition (\ref{res}) in term of conservation of energy an momentum on a microscopic level, and this is helpful in interpreting the anomalous Doppler resonance. In a (relativistic) quantum treatment the energy states of an electron are $\varepsilon=\varepsilon_n(p_\parallel)=(m^2c^4+p_\parallel^2c^2+p_\perp^2c^2)^{1/2}$, $p_\perp^2=2\hbar neB$, with $n\ge0$ the Landau quantum number. The transition assumed is $\varepsilon\to\varepsilon-\hbar\omega$, $p_\parallel\to p_\parallel-\hbar k_\parallel$, $n\to n-s$. The gyroresonance condition (\ref{res}) follows from conservation of energy and momentum in the limit $\hbar\to0$. The anomalous Doppler transition here is for an electron initially in the Landau ground state, $n=0$, to its first excited state $n=1$, so that the perpendicular energy increases, despite  energy being carried away by the emitted photon. This is possible only for subluminal phase speeds. Assuming that the refractive index is $1+\Delta n$ and that the angle, $\theta$, of emission is small, the frequency implied by (\ref{res}) becomes
\be
\omega\approx{2\gamma\Omega_\rme\over1+2\gamma^2\Delta n+\gamma^2\theta^2}\approx{2\Omega_\rme\beta_A^2\over\gamma},
\label{res1}
\ee
where the final approximate relation applies to the x~mode, cf.\ (\ref{infB2}). The frequency (\ref{res1}) is well above the radio range near the surface of a neutron star. 

The anomalous Doppler instability is relevant for radio emission only if it develops far from that stellar surface, where the frequency (\ref{res1}) is in the radio range. The strong dependence of this frequency on $B$ seemingly excludes anomalous Doppler emission being the common radio emission mechanism for all three classes of pulsars.

\section{Discussion and conclusions}
\label{sect:discussion}

After nearly half a century of research on pulsars, we have no consistent global model for pulsar electrodynamics, and we have no widely accepted pulsar emission mechanism.  Differences of opinion on the electrodynamics and on the emission mechanism remain, and some of these have developed into long-standing misconceptions. Notable examples are the aligned assumption in the electrodynamics, and coherent curvature emission by bunches as the emission mechanism. 

The aligned assumption, and the related electrostatic assumption in an obliquely corotating frame \citep{FAS77,SAF78}, are important from the practical viewpoint of allowing detailed models to be developed, but any realistic model must be based on an oblique rotator. An aligned model leads to the neglect of ${\bi E}_{\rm ind}$ and ${\bi J}_{\rm disp}$, thereby excluding two effect that we identify here. First, inclusion of ${\bi E}_{\rm ind}$ allows an azimuthal dependence to develop across a gap, and this leads to a possible new model for subpulses, \S\ref{sect:subpulses}.  Second, inclusion of ${\bi J}_{\rm disp}$, in a corotation model, leads to the inevitability of current starvation, \S\ref{sect:current_starvation}, with possibly important implications.  

Similarly, coherent curvature emission by bunches may seem to have the advantage of simplification, in that the properties of the emission, notably its frequency spectrum and polarization, can be assumed to be those of single-particle curvature emission, enabling comparison of the observational data with a known theory. However, this is the case only for a point-like bunch. For a more realistic bunch, the frequency and angular dependences are modified by a factor $|{\tilde n}({\bi k})|^2$, as discussed in \S\ref{sect:curv}. Moreover, in the absence of a plausible mechanism for creating the postulated bunches, coherent curvature emission should not be regarded as a realistic mechanism.  

An essential ingredient in any realistic model for a pulsar magnetosphere is the need for regions with $E_\parallel\ne0$, cf.\ \S\ref{sect:need}. Such regions are  `gaps' in RMMs and current sheets in FFE models. However, the location and properties of such gaps are poorly constrained by theory, and this has led to misconceptions. An older misconception is that gaps are electrostatic structures within which particle acceleration occurs. Effective particle acceleration by electrostatic structures is known to be impossible in principle \citep{Betal92}.  Effective acceleration must be due to incomplete screening of $E_{\rm ind\parallel}$, due to charge (or current) starvation. This misconception does not necessarily lead to significant error due to models for gaps being 1D. There is no distinction between inductive and potential fields in a 1D model, so that there is no quantitative error, provided that the value of the potential drop in an electrostatic model is based on the inductive counterpart. Static models for gaps are violently unstable to temporal perturbations, and gaps need to be re-interpreted in terms of propagating wave-like structures with $E_\parallel\ne0$, as discussed in \S\ref{sect:LALOs}. A potentially more serious misconception, emphasized by  \cite{SL06}, relating to an appeal to a generalized Ohm's law discussing the accelerating electric field. It is essential that the accelerating field be attributed the electrodynamics, specifically to ${\bi E}_{\rm ind}$ and ${\bi J}_{\rm disp}$.  The intrinsic time dependence of acceleration and pair creation are essential ingredients, which are included in some detailed numerical models \citep{T10,TA13}. 

In summary, while much progress has been made in understanding pulsars, there remain major unsolved problems in pulsar electrodynamics, and there is no consensus of the specific radio emission mechanism. We need to understand the electrodynamics to identify the source of the radio emission, and we need to identify the radio emission mechanism in order to use the radio data to constrain the properties of the pulsar magnetosphere. Progress towards these objective is frustratingly slow.

\section*{Acknowledgments} 
RY acknowledges supports from Project 11573059 NSFC; and the Technology Foundation for Selected Overseas Chinese Scholar, Ministry of Personnel of China.

\bibliographystyle{jpp}

\bibliography{jpp-instructions}

\end{document}